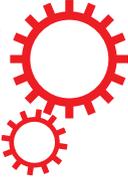

**OPEN**

# Blue mussel shell shape plasticity and natural environments: a quantitative approach

Luca Telesca[1,2], Kati Michalek[3], Trystan Sanders[4], Lloyd S. Peck[2], Jakob Thyrring[5,6] & Elizabeth M. Harper[1]

Shape variability represents an important direct response of organisms to selective environments. Here, we use a combination of geometric morphometrics and generalised additive mixed models (GAMMs) to identify spatial patterns of natural shell shape variation in the North Atlantic and Arctic blue mussels, *Mytilus edulis* and *M. trossulus*, with environmental gradients of temperature, salinity and food availability across 3980 km of coastlines. New statistical methods and multiple study systems at various geographical scales allowed the uncoupling of the developmental and genetic contributions to shell shape and made it possible to identify general relationships between blue mussel shape variation and environment that are independent of age and species influences. We find salinity had the strongest effect on the latitudinal patterns of *Mytilus* shape, producing shells that were more elongated, narrower and with more parallel dorsoventral margins at lower salinities. Temperature and food supply, however, were the main drivers of mussel shape heterogeneity. Our findings revealed similar shell shape responses in *Mytilus* to less favourable environmental conditions across the different geographical scales analysed. Our results show how shell shape plasticity represents a powerful indicator to understand the alterations of blue mussel communities in rapidly changing environments.

Exploring shape variability and uncovering its underlying causes is essential to understand the diversity of life, as well as to appreciate the great heterogeneity of forms that exist in nature[1–3]. Physical constraints are of primary importance in determining the form of an organism as minor variations in growth processes can lead to dramatic shape alterations[1,4]. Therefore, developing rigorous methods to quantify shapes and describe their natural variation could provide a better understanding of the underlying mechanisms driving the diversity of biological forms.

Bivalves constitute a substantial component of coastal benthic communities[5]. Among them, blue mussels, *Mytilus* spp., are important foundation species throughout the temperate and polar littoral zones of the northern and southern hemispheres[6,7], and represent an important economic resource for the aquaculture industry[8].

A number of studies have shown a variable distribution of blue mussel species at a North Atlantic scale[6,9]. In the *Mytilus edulis* species-complex (*Mytilus edulis*, *M. trossulus* and *M. galloprovincialis*), an extensive hybridisation pattern has been documented wherever the ranges of these three species overlap[6,10] and reviewed in an aquaculture context[11]. This has a potentially complicated influence on mussel shapes[12,13].

During the last two decades, much attention has been paid to climate change and its evident effects on calcifying marine organisms[14,15]. Heterogeneous patterns of environmental variation and increasing anthropogenic pressures have highlighted limitations in our ability to forecast emergent ecological consequences of global changes[16]. There is, therefore, a clear need for knowledge on the processes regulating marine ecosystems and their resilience[17]. These issues are creating new challenges for understanding organismal responses to key environmental drivers, which is essential for predicting sensitivity to multiple stressors and improving our ability to forecast alterations at higher levels of organisation[17].

Atlantic *Mytilus* spp. have been widely used as model organisms for studying ecological and physiological responses to different environmental conditions[18–20]. Growing awareness of climate change and its consequences for the considerable biodiversity that blue mussels support[21,22] have sparked interest in predicting sensitivity of these habitat-forming species[22,23]. Indeed, the understanding of the significance of morphological variation in *Mytilus*[24,25] is increasing in parallel with the development of statistical tools to predict species-specific responses[26].

[1]Department of Earth Sciences, University of Cambridge, CB2 3EQ, Cambridge, United Kingdom. [2]British Antarctic Survey, CB3 0ET, Cambridge, United Kingdom. [3]Scottish Association for Marine Science, PA37 1QA, Oban, United Kingdom. [4]GEOMAR Helmholtz Centre for Ocean Research, 24148, Kiel, Germany. [5]Department of Bioscience, Arctic Research Centre, Aarhus University, 8000, Aarhus C, Denmark. [6]Department of Bioscience, Marine Ecology, Aarhus University, 8600, Silkeborg, Denmark. Kati Michalek and Trystan Sanders contributed equally to this work. Correspondence and requests for materials should be addressed to L.T. (email: lt401@cam.ac.uk) or E.M.H. (email: emh21@cam.ac.uk)

Growth and shape of mussels and the degree to which they vary with respect to environmental factors have been documented for numerous species and habitats[23,26,27]. In fact, *Mytilus* shell changes can reflect responses to conditions selecting for specific traits[24,27,28] and the level of shape variation may be used as a good indicator of habitat change. Documented shell modifications under forecasted conditions could potentially increase mussel sensitivity to biotic and abiotic drivers[23,24] and have profound indirect impacts on this foundation species with cascade consequences for supported communities and ecosystems[22,29,30]. Therefore, multi-population studies across broad geographical areas, spanning a range of environmental conditions, are critical to identify organismal responses to drivers in a multivariate natural environment[17,23].

A range of qualitative[28] and quantitative[12,24,26] methods have been used to describe the variation in shell traits (morphometrics) and outline (shape) of *Mytilus* in relation to environment and genotype. Standard approaches constitute traditional morphometrics and regression-type analyses[13,31]. However, their application can result in predictions with poor accuracy of the factors driving shell shape[32] and have implications for the understanding of plasticity.

Traditional morphometrics, which involves applying multivariate analysis to sets of linear descriptors, can mask phenotypic responses[31]. Indeed, mussels can be characterised by variations in shell features that are difficult to quantify (e.g. umbo orientation, convexity of the ventral margin)[33] showing fine-scale shape patterns without alterations of linear shell dimensions. In contrast, the development of geometric morphometrics has emphasised the potential to capture the geometry of the features of interest[2] and to provide powerful analyses of bivalve shape variation[12,34,35]. Unlike ordinary least square methods, newly developed generalized additive mixed models (GAMMs)[36] offer ways to account for the hierarchical structure of ecological datasets, and are powerful tools for defining flexible dependence structures as well as dealing with heterogeneous distributions[37]. However, a combination of these methods and their inferential advantages have never been applied to heterogeneous patterns of organismal shapes in natural environments.

The aims of this study are to (**i**) quantify shell shape variation in North Atlantic and Arctic *Mytilus* species from different geographical regions through an elliptic Fourier analysis (EFA) of outlines, (**ii**) identify general and local environmental effects on shell shape mean and heterogeneity, through the use of GAMMs and study systems at various geographical scales, (**iii**) show how the use of new methods allows the uncoupling of environmental, developmental (age) and genetic (species) contributions to *Mytilus* shape and the description of relationships between blue mussel shape variation and environment that are independent of age and species influences, and (**iv**) test the hypotheses that environmental covariates drive the among-individual shell shape variation and environmental stressors can induce the formations of similar shapes at the different geographical scales of analysis. This work further aims to reveal previously unrecognised fine-scale shape responses in Atlantic blue mussels and to estimate effect sizes of different drivers on shape variation.

By providing a representative sample for the distribution of blue mussels as well as powerful methods to identify factors influencing shell shape plasticity, it would become possible to appreciate the great variation in mussel forms that exist in nature[28].

## Material and Methods

**Mussel collection.** We sampled a total of 16 *Mytilus* spp. populations living along the North Atlantic, Arctic and Baltic Sea coastlines from three study systems at different geographical scales (large-, medium- and small-scale).

Shell shape variation among habitats was analysed across a large geographical scale, System 1, on ten wild blue mussel populations (sites 1–10; Fig. 1a) sampled at different latitudes from four distinct climatic regions (warm temperate, cold temperate, subpolar and polar). Mussel specimens were collected from Western European (Exmouth, England, 50°N) to Northern Greenlandic (Qaanaaq, 78°N) coastlines, covering a latitudinal range of 28° (a distance of 3980 km). Environmental influence on a medium spatial scale, System 2, was investigated using five wild mussel populations (sites A-E; Fig. 1b) collected from the North Sea (Sylt, Germany) to the innermost part of the Baltic Sea (Nynäshamn, Sweden). In addition, we studied shell shape variation on a small geographical scale, System 3, using specimens obtained from a traditional longline mussel farm on the Scottish west coast (Loch Leven, UK; Fig. 1c). Four batches of mussels, originating from a natural spatfall, were collected at one, three, five and seven metres depth (batches I, III, V and VII, respectively), representing the natural distribution of mussels along the cultivation ropes.

During December 2014 and January 2016, we collected mussels of various size classes for each population (shell length 25–81 mm) for a total of 555 individuals. Wild adult mussels (System 1 and 2) were sampled from the eulittoral zone and cultured specimens (System 3) were harvested as part of a long-term monitoring programme (Supplementary Table S1). For each specimen, we measured shell dimensions with a digital calliper (0.01 mm precision) (Supplementary Figure S1a,b), among which shell length was used as a within-population proxy for age[5,27,38].

We examined *Mytilus* populations with available information on their genotype, with a particular focus on species identity and documented hybridisation (*Mytilus edulis*, *M. trossulus* and *M. edulis* × *M. trossulus* hybrids). Blue mussels used were from populations recently analysed in genetic investigations, sites routinely employed in regional monitoring programmes and specimens already used for genetic analyses (Supplementary Table S1). Therefore, we used populations with a known genetic status.

Reference populations of *Mytilus edulis* and *M. trossulus* were selected from two sites in western Europe, one site in Greenland and one Baltic location (populations 1, 4, 10 and E, respectively; Fig. 1a,b). According to genetic analyses of these populations, which are based on multiple genetic markers (multi-locus genotyping) or SNP analyses, these samples are representative of these two species[7,39,40]. Although molecular studies have revealed various episodes of introgression and hybridisation, which increases the evidence that no completely pure reference populations exist in the North Atlantic and Baltic Sea[10,12], our reference samples are as representative as possible considering the geographical range of the study. Therefore, these provide a solid starting point for the following among-species shell shape comparisons. Given the absence of representative populations of *M. galloprovincialis* at the analysed spatial scale, we avoided areas where this species was either present or there was a high degree of hybridisation (e.g. south-central Norway, parts of continental European and Ireland's coastlines)[41,42]. We did, however, sample sites where very low proportions of *M. edulis* × *M. galloprovincialis* hybrids have been reported[9].

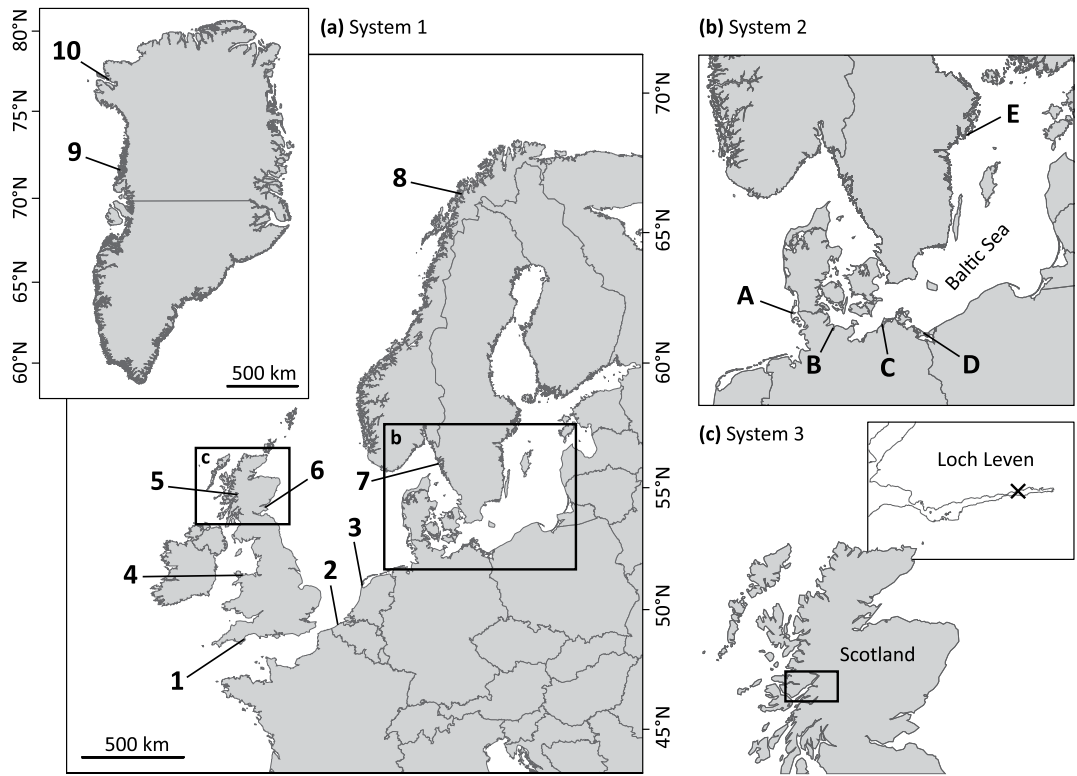

**Figure 1.** Blue mussels collection sites. Study systems and locations where *Mytilus* populations were collected. (**a**) System 1, large-scale North Atlantic and Arctic regions: (1) Exmouth, south-west England, (2) Oostende, Belgium, (3) Texel, north Netherlands, (4) Menai Bridge, north Wales, (5) Tarbet, west Scotland, (6) St. Andrews, east Scotland, (7) Kristineberg, west Sweden, (8) Tromsø, north Norway, (9) Upernavik and (10) Qaanaaq, west Greenland. (**b**) System 2, medium-scale Baltic region: (A) Sylt, (B) Kiel, (C) Ahrenshoop, (D) Usedom, all Germany and (E) Nynäshamn, east Sweden. (**c**) System 3, small-scale: (X) long-line mussel farm (Glencoe Shellfish Ltd.) in Loch Leven, west Scotland, with four sampling depths (I, III, V and VII meters). Map created with ArcMap 10.3 (ArcGIS software by Esri, http://www.esri.com) using data from ©OpenStreetMap (http://www.openstreetmap.org).

**Environmental parameters.** We selected environmental covariates according to the availability of data for the investigated areas and their known effects on growth, development and mussel energy budgets[23,27,43]. Given the high collinearity of many physical and biogeochemical descriptors at the geographical scale considered, we chose three key parameters: water temperature, salinity and chlorophyll-a (chl-a) concentration, the latter being validated as a proxy for food availability[18,26]. Predictors for the large- and medium-scale systems (System 1 and 2) were generated using the Copernicus Marine Environment Monitoring Service (CMEMS)[44]. These datasets are composed of high-resolution physical and biogeochemical analyses of assimilated (integration of observational and predicted information) daily data (Supplementary Document S1). For each parameter, mean values per site for the 2014–2015 period were used as predictors. For the large- and medium-scale systems, remote-sensing and assimilated data presented potential advantages compared to traditional measurements[26,45] due to their known high spatial and temporal resolution, advanced calibration and validation (i.e. high correlation with discrete field measurements)[44,46]. Environmental parameters for the small-scale system (System 3) were calculated from samples collected fortnightly over the course of a year and expressed as annual mean values for each depth (Michalek, *pers. obs.*; Supplementary Document S1).

**Elliptic Fourier analysis of shell outlines.** Shape analyses of *Mytilus* shells were performed through a geometric morphometrics approach[2]. An elliptic Fourier analysis (EFA) of outlines[4,47] was used to examine shell shape variation both within and between populations from different study systems. The improved EFA method presents several advantages compared to older approaches[4]: complex shapes can be fitted, outlines smoothed, starting points normalised, and homothetic, translational and rotational differences removed[4,34,48,49]. Shape analyses were carried out using the package Momocs[4] in the software R[50].

Outlines of orthogonal lateral and ventral views of the left valves were digitised, converted into a list of *x-y* pixel coordinates and used as input data. The outlines for both views were then processed independently, geometrically aligned and later combined for analysis (Supplementary Methods and Figure S1c). We then computed an EFA on the resulting coordinates from shapes invariant to outline size and rotation. After preliminary calibration, we chose seven harmonics, encompassing 98% of the total harmonic power[34]. Four coefficients per harmonic were extracted for each shell outline and used as variables quantifying the geometric information[48].

A principal component analysis (PCA), with a singular value decomposition method, was performed on the matrix of coefficients to observe shape variation among individuals and populations from the different study systems. Calculated principal components (PCs) were considered as new shape variables. To understand the contribution of individual variables to shell shape, we reconstructed extreme outlines along each PC. The first 10 PCs, accounting for 97% of outline variation, were analysed with a multivariate analysis of variance (MANOVA) to test for significant effects of the location of origin and shell length (size) on shape variances. To visualise differences at the extremes of the morphospace, we generated deformation grids[1] and iso-deformation lines through mathematical formalisation of thin plate splines (TPS) analysis[51].

For the reference populations of *M. edulis* and *M. trossulus*, we performed a linear discriminant analysis (LDA) based on the new shape variables (PCs), with a leave-one-out cross-validation procedure, to identify the linear combination of shape features that was able to discriminate between *Mytilus* species. Standardised coefficients from the calculated discriminant function were used to compare the relative importance of each shape variables (PCs) at discriminating between species. We set *a priori* classification probabilities to be proportional to group sizes and Wilks' λ were calculated to test for significant discrimination. Discriminant coefficients were estimated to identify shell shape features that optimised the between-species differences "relative" to the within-species variation[48].

**Data exploration and statistical modelling.** The first five PCs, capturing 91% of shape variance and describing distinguishable features along the outline, were selected for analysis.

Generalised additive mixed models (GAMMs)[36,37] were used to explain shape variance with respect to mean environmental parameters and shell size, and to compare between individual shape features (PCs). A GAMM is a generalised linear mixed model with a linear predictor involving a sum of smooth functions of covariates[36]. The model allowed for flexible specification of the dependence of the response on the covariates by defining regression splines (smoothing functions) and estimating their optimal degree of smoothness, rather than calculating parametric relationships[36].

Given the dependency on the same set of predictors, we analysed all the PCs for each study system within the same model. This new approach allowed accounting for the dependence of multiple shape variables, which describe synergistically the shell outline as a whole (as implied by the adopted EFA method), and defining combinations of linear and non-linear relationships simultaneously. The among-individual shape variances were then analysed together without losing descriptive power or increasing the probability of Type I error[32]. We performed all data exploration and statistical analyses in R[50]. Models were fitted using the mgcv[36] and nlme[52] packages.

Initial data exploration, following the protocol of Zuur *et al.*[53], revealed no outliers. Conditional boxplots showed heterogeneous shape variances (PCs eigenvalues) as a procedural consequence of the PCA. This required standardisation prior to analysis since we were not interested in between-feature heterogeneity[32]. Response variables did not require any transformation. Pairwise scatterplots and calculation of variance inflation factors (VIFs)[53] indicated low collinearity between predictors for System 1 and 2. For these systems, the effects of multiple environmental covariates on shape variance were modelled simultaneously only if VIFs < 3[53]. We detected high collinearity among environmental predictors for System 3. Therefore, we performed a PCA on these explanatory variables to calculate new linear combinations of covariates accounting for the greatest variation in the original values[54]. We then used scores of orthogonal PCs (enviro-PC1, 2 and 3) as new independent environmental predictors to model shape variance. In addition, potential interactions between continuous covariates and shape features, and clear non-linear patterns were detected.

We used GAMMs to model shape variance for (i) large-scale (System 1), (ii) medium-scale (System 2), (iii) small-scale (System 3) study systems and (iv) the pooled mussel populations (Atlantic system; Equation [1]). We employed a combination of a single question approach (individual systems) and an analysis of the pooled populations (Atlantic system) to model and differentiate local environmental effects, being more dependent on the geographical scales considered, from the general effects of environmental variation, having a more consistent influence on the shell shape of blue mussels from different regions.

To model shape variance as a function of environmental covariates for the Atlantic system, we used a GAMM with a normal distribution (Equation [1]). Fixed continuous covariates used were water *temperature*, *salinity* and *chl-a* concentration all fitted as smoothers, in addition to *shell length* (continuous), shape features (*PC*, categorical with five levels) and their interactions with continuous predictors. To incorporate the dependency among specimens from the same site of collection, we used *site* as a random effect. The final model was of the form:

$$
\begin{aligned}
ShapeVar_{ijk} &\sim N(\mu_{ijk}; \sigma^2) \\
\mu_{ijk} &= f(Temperature_i) \times PC_j + f(Salinity_i) \times PC_j + f(Chl-a_i) \times PC_j \\
&\quad + Length_{ik} + PC_j + Length_{ik} \times PC_j + Site_i \\
Site_i &\sim N(0, \sigma^2_{Site})
\end{aligned}
\tag{1}
$$

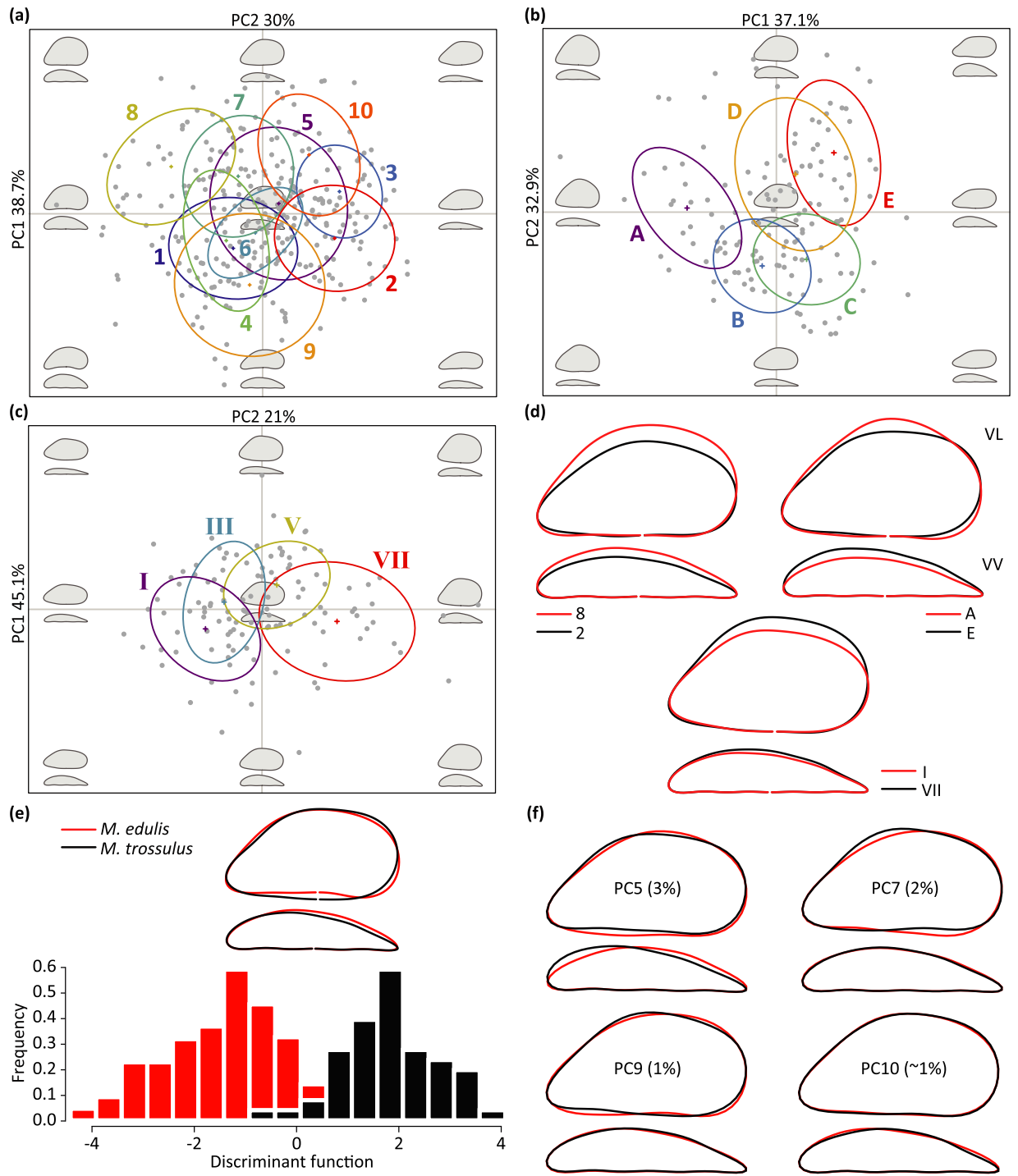

**Figure 2.** Among-population variations in outlines and shape features. Scatterplots of the first two PCs from PCAs performed on the elliptic Fourier coefficients, of lateral and ventral shell views, showing a clear separation and marked shape variation among specimens from (**a**) System 1 (large-scale), (**b**) System 2 (medium-scale) and (**c**) System 3 (small-scale). Confidence intervals for each group of origin and the reconstructed morphospace (background) are shown. (**d**) Mean shape differences of lateral (VL) and ventral (VV) views between populations or batches at the extremes of the morphospace: System 1 populations 8–2, System 2 locations A - E, and System 3 batches I - VII. Population 8 had rounder and wider shells, with higher and more convex ventral sides than population 2 (elongated and narrow shells). Location A was characterised by rounder and higher shells with a more convex ventral side than site E (elongated and wide shells with almost parallel dorsoventral margins). Mussels from batch I displayed more elongated shells, with a smaller height, ligament area and width than round mussel specimens from batch VII. All the reconstructed mean outlines showed a consistent variation in distinct shell features: shell height, ventral side shape, ligament area and shell width. (**e**) Discriminant function calculated from a LDA on the new shape variables showing a significant separation between species (Wilk's $\lambda = 0.264$, approx. $F_{1,113} = 29.00$, $p < 0.0001$) and differences between the mean shell outlines for the individual groups (red: *Mytilus edulis*; black: *M. trossulus*). (**f**) Shape variables (PC5, 7, 9 and 10) contributing

the most to discriminate between species. These PCs captured the species-contribution to the shell shape of *Mytilus* (7% of total shape variance). Individual contributions to the mean outline are represented through the reconstruction of mean shapes for high (red) and low (black) PCs values (Mean ± 3 SD, respectively).

where $ShapeVar_{ijk}$ is the $k^{th}$ observation for $j^{th}$ PC ($j = 1, ..., 5$ levels) and $i^{th}$ site ($i = 1, ..., 15$ levels). $f$ is the smoothing function and $Site_i$ is the random intercept, which is assumed to be normally distributed with mean 0 and variance $\sigma^2_{Site}$. The $f(continuous\ predictor) \times PC$ interaction applies a smoother on the data for each PC.

We manually selected the optimal amount of smoothing and a cubic regression spline was applied[36]. Variograms indicated no spatial or temporal autocorrelation. Statements about trends of shape variance and environmental gradients are based on the significance (at $\alpha = 0.01$) of individual interaction terms between predictors and PCs. Models were optimised by first selecting the random structure and then the optimal fixed component[32,55]. Visual inspection of residual plots indicated a violation of homogeneity in most cases. This required the use of specific variance structures (generalised least squares) allowing the residual spread to vary with respect to continuous predictors and shape features[32]. Once we found the optimal model (in terms of the random structure), we applied further selection by rejecting any non-significant interaction term between the explanatory variables. The principal tools for model comparisons were the corrected Akaike Information Criterion (AICc) and likelihood ratio tests for each nested model. Final models (Supplementary Table S2) were validated by inspection of standardised residual patterns to verify the assumptions of normality, homogeneity and independence[32]. We used models to predict trends with environmental gradients and estimate the mean effect sizes (same measurement unit) of standardised environmental parameters. For standardisation, we subtracted the sample mean from the variable values and divided them by the sample standard deviation. Confidence intervals (95%CI) and mean effect sizes were estimated to compare the magnitude of the effect of individual covariates on the responses. If the confidence intervals did not overlap with zero, the effect size was considered significant.

## Results

**Mussel geometric morphometrics.** The first two PCs, from PCAs performed on harmonic coefficients, accounted for 68.7%, 70.0% and 66.1% of the shape variation among individuals from Systems 1, 2 and 3 respectively, and 70.2% of variance for the Atlantic system. Scatterplots of PC1 and PC2 showed a clear separation among groups across the morphospace (Fig. 2a–c, Supplementary Figure S2a), revealing marked among-individual variation for both lateral and ventral views.

For the Atlantic system, MANOVAs revealed significant effects of collection site (Wilk's $\lambda = 0.032$, approx. $F_{1,419} = 12.61$, $p < 0.0001$) and shell length (Wilk's $\lambda = 0.873$, approx. $F_{1,419} = 5.95$, $p < 0.0001$) on shape variance. Additionally, significant influences of location of origin and shell length at different geographical scales were identified (Supplementary Table S3). Mean shapes and TPS analyses indicated the main outline deformations required to pass from one extreme of the morphospace to another (Fig. 2d, Supplementary Figures S2b, S3).

We identified shape features that contributed the most to the observed patterns of shape variation for different systems through comparison of reconstructed outlines at the extremes of the morphospace along each axis. The first five PCs, depicting the variation in specific shell features, were described through their individual contribution to the outline reconstruction for increasing PC values, for individual study systems (Supplementary Table S4 and Figures S4, S5, S6) and the Atlantic system (Table 1, Supplementary Figure S7).

A LDA based on the new shape variables allowed us to identify the shape features that discriminate most between the two *Mytilus* species and to isolate the species-contribution to the shape variance. Ninety-seven percent of individuals were correctly reclassified by the new discriminant function (Fig. 2e). The LDA on the reference populations produced an efficient separation between groups and a cross-validation (leave-one-out) at species level showed a high confidence in the reclassification (98.3% and 94.6% of correct reattribution for *M. edulis* and *M. trossulus*, respectively). Standardised discriminant coefficients indicated PC5 (3%), PC7 (2%), PC9 (1%) and PC10 (~1%) had the highest contribution to the separation between species. We selected these PCs as the variables capturing the most of the shape information on the species-contribution to *Mytilus* shell shape. The identified variables contributed to subtle variations in shell outlines (Fig. 2f) and showed limited overlap with the shape features described by the PCs capturing the most of shape variance among individuals (PC1-PC4).

**Shell shape variation and environmental factors.** GAMMs indicated highly significant relationships between the axes capturing most of the shape variation and environmental parameters for all the study systems, with associations depending on the shape features (PCs) considered (Fig. 3). Only significant relationships ($p < 0.01$) are presented in the following section.

*System 1 – Large geographical scale.* The model (Table 2) showed positive and negative non-linear relationships of PC1 with temperature and salinity, respectively. We detected associations of PC2 with salinity, shell length and chl-a. PC3 and PC4 showed marginal negative relationships with temperature and chl-a, respectively. Overall, we observed the formation of elongated and narrow shells with decreasing temperature and salinity (Fig. 4a), and a transition from elliptical to more elongated, curved and wider profiles with increasing food supply and shell length (Supplementary Figure S8). An exponential variance structure indicated a negative effect of water temperature ($df = 5$, $L = 39.82$, $p < 0.0001$) on shape variance.

| PCs | %Variance | Contribution to shell shape |
|---|---|---|
| PC1 | 38.1 | Shell height, the shape of ventral and dorsal margins, ligament length and shell width: low values corresponded to elliptical shells, with concave ventral margins, long ligaments and narrow profiles, while high values were associated with "curved", wide shells, with convex ventral margins and short ligaments. |
| PC2 | 32.1 | Shell height, ligament angle and width, describing a gradual transition from round and wide shells to elongated and narrow mussels for increasing values. |
| PC3 | 11.6 | Shape of ventral margin, umbo and ligament with small variations of shell width: negative values corresponded to shells with concave ventral margins, convex ligaments and an umbo oriented toward the anterior side, while positive values to more "curved" shells with concave ventral margins, convex ligaments and an umbo pointing downwards. |
| PC4 | 5.1 | Contribution to the shape of ventral margin and variability between "curved" (concave ventral margins) and elliptical shells. |
| PC5 | 3.0 | Small variations in dorsoventral shape (more or less parallel margins) and the symmetry of ventral view. |

**Table 1.** PCs contribution to the shell shape. Proportion of shape variance captured by individual shape variables and description of their contribution to the shell features and mean shape reconstruction (Supplementary Figure S4).

*System 2 – Medium geographical scale.* GAMMs (Table 2) indicated a non-linear association between PC1 and salinity only. PC2 and PC5 showed negative relationships with salinity and temperature. The model revealed a general positive effect of shell length ($df = 1$, $F_{1,126} = 7.75$, $p = 0.0055$). Overall, we found more elongated, wide shells and more squared margins with decreasing salinity and temperature (Supplementary Figure S9a). Round mussels with big ligament areas were associated with high salinities (~30 psu), while elongated, wide shells were identified in low salinities (~6 psu) (Fig. 4b). An exponential variance structure indicated a positive effect of chl-a concentration ($df = 5$, $L = 52.05$, $p < 0.0001$) on shape variance.

*System 3 – Small geographical scale.* Model selection reported significant effects of enviro-PC1 only along the cultivation rope (Table 2). PCA indicated an equal positive contribution of water temperature and salinity, and a negative contribution of chl-a concentration to the enviro-PC1 loadings. The optimal model showed a positive non-linear relationship with PC1, a marginal increasing trend with PC2 and a non-linear association with PC5 (Supplementary Figure S9b). Our results indicated a progressive increase in shell height, width and ligament length with increasing values of enviro-PC1, showing a transition from elongated and narrow to round and wide mussel shells with increasing temperature, salinity, and decreasing food availability (Fig. 4c).

*Atlantic system.* Equation 1 indicated relationships between blue mussel shape and all the modelled predictors (Table 3, Fig. 3a). PC1 showed positive relationships with temperature and shell length, and non-linear patterns with salinity and food availability. PC2 indicated non-linear relationships with temperature and chl-a, a negative association with salinity and a positive one with shell length. We detected positive associations of PC3 with both temperature and salinity. PC4 was characterised by a positive relationship with temperature and non-linear association with salinity and food availability. Overall, we identified the formation of elongated, narrow shells and more squared margins with decreasing salinity, an increasing shell height and width with increasing chl-a and a transition from elliptical to elongated, curved and wider shells with increasing temperature and shell length (Fig. 3a). We specified exponential variance structures[32] allowing residuals to vary with respect to surface temperature and shape features (PCs). The best variance structure indicated a negative exponential effect of temperature ($df = 5$, $L = 59.65$, $p < 0.0001$) on shape variance.

Mean effect sizes revealed differences in the relative contribution of modelled covariates (Table 3, Fig. 3b). PC1 was characterised by a marked effect of shell length and environmental influences of temperature and chl-a. Water salinity had the strongest effect on PC2, being about three times bigger than the effect of shell length. We also found a marked influence of temperature, salinity and a weak effect of length on PC3, while PC4 was strongly influenced by all the environmental descriptors. We detected no effect on PC5.

## Discussion

Shape analysis is a fundamental component of several areas of biological research[2]. In ecology, it can allow discrimination of shapes of organisms from specific habitats and understanding of the underlying mechanisms leading to variation of morphological structures[4]. This is especially important for economically and ecologically valuable taxa, such as blue mussels[6,20]. With regards to aquaculture, shape variations under changing environments could produce fragile shelled mussels[24]. These are less valuable economically[56], being easily damaged during harvest, grading and transport processes, and may lead to significant financial losses for the industry[57]. In natural habitats, changes in shapes and structural integrity of shells can increase their vulnerability to predation[24,58], with potential profound impacts on whole ecosystems[22].

Geographical variation in *Mytilus* shell shape is confounded by marked shell modifications during growth[28] and among-species differences[12], on top of which environmental heterogeneity strongly influences spatial shape patterns[27]. Several studies have explored the effects of these individual factors on natural shape variations in different mussel species. Seed[28,38] investigated the influence of growth rate and age on *M. edulis* form, providing a qualitative baseline for the interpretation of its developmental changes. The effect of genotype on shell shape and morphology has also been explored for the Atlantic mussels, showing differences among taxa in various geographical regions[12,13,57]. Modelling was used to identify the relationships between mussel growth and environmental factors across relatively small spatial scales[26], while broad-scale studies have highlighted consistent morphometric patterns along latitudinal gradients in the South Pacific[33,35]. In addition, experimental-induced phenotypic shape responses have indicated potential deleterious effects of future increases in temperature and pCO$_2$ on shell integrity of *M. edulis*[24]. Moreover, a body of research showed many factors have more local influences on shape, such as hydrodynamic regimes, ice cover and parasitic diseases[6,13,59]. However, our ability to forecast heterogeneous patterns of mussel shape responses to altered environmental conditions in multi-population studies is limited by our ability to uncouple the contributions of developmental (age) and genetic (species) factors from shell shape variations. Specifically, heterogeneous size classes and multiple species prevent us from identifying general relationships between *Mytilus* shape variation and local drivers without selectively controlling for these two confounding factors (i.e. analysing similarly sized individuals and/or different species separately).

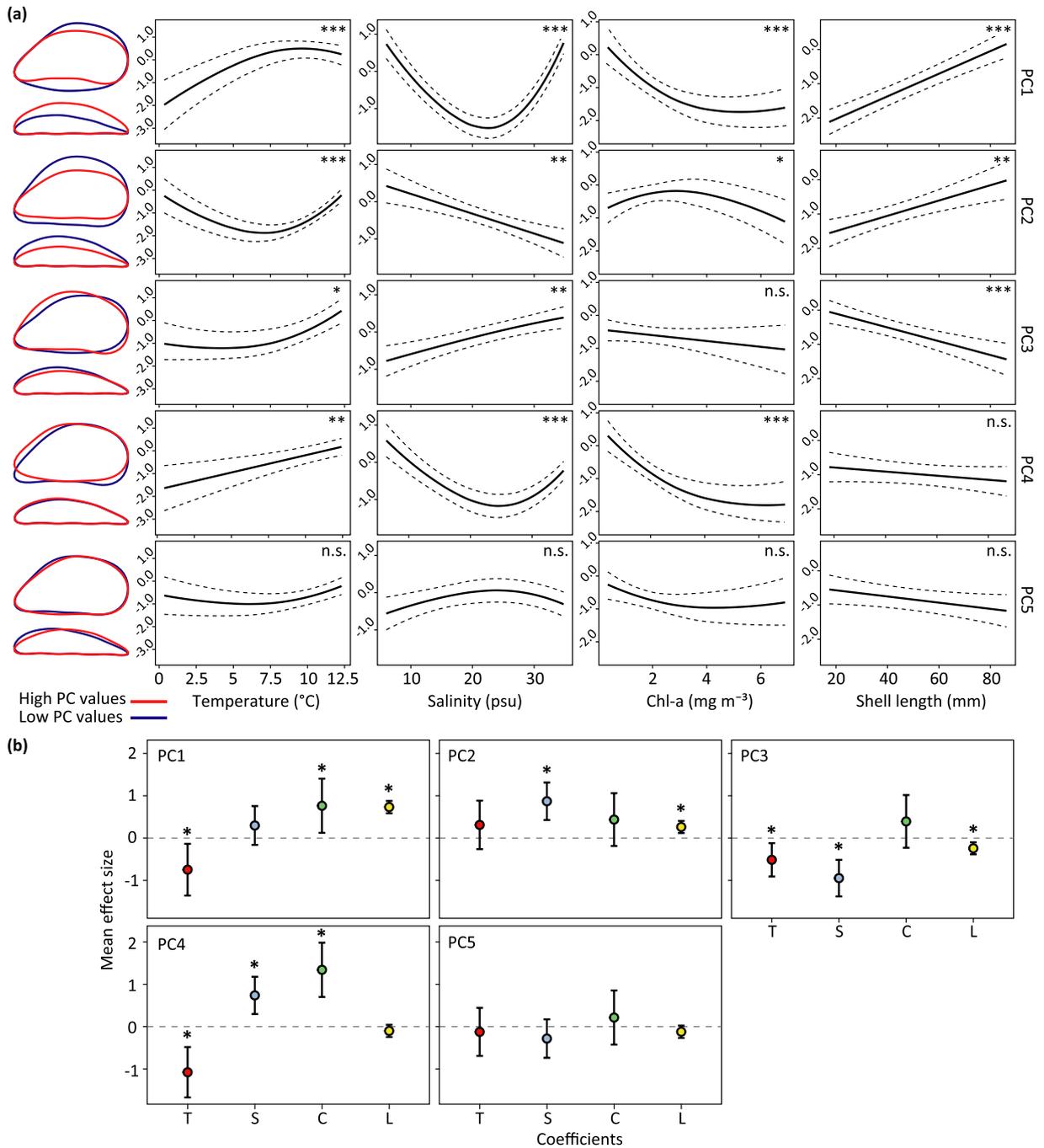

**Figure 3.** *Mytilus* shell shape patterns and effect sizes. (**a**) Modelled shape trends of individual shell features (PC1–5) with environmental descriptors (sea surface temperature, salinity and chl-a concentration) and shell length (size) for the mussel populations from the Atlantic system (Equation 1). Predicted values (continuous lines), 95%CI (dashed lines) and significance level (accuracy of estimated standard errors) of each fitted smoother and linear predictor are shown. Mussel shape variations described by each shape variable are represented through the comparison of mean outlines reconstructed for low and high PC values (blue: Mean − 3 SD; red: Mean + 3 SD). (Significance, n.s. $p > 0.01$, *$p < 0.01$, **$p < 0.001$, ***$p < 0.0001$). (**b**) Mean effect sizes of temperature [T, Mean (SD) = 8.87 °C (5.55)], salinity [S, Mean (SD) = 26.15 psu (10.54)], chl-a [C, Mean (SD) = 2.13 mg m$^{-3}$ (1.50)], shell length [L, Mean (SD) = 50.35 mm (17.03)] for individual shape variables (PCs) and their significance. Error bars represent 95%CIs. Significance is determined when the confidence interval does not cross zero (*$p < 0.01$).

| System | PCs | f(Temperature) × PCs | | | f(Salinity) × PCs | | | Chl-a × PCs | | | Length × PCs | | |
|---|---|---|---|---|---|---|---|---|---|---|---|---|---|
| | | edf | F | p | edf | F | p | df | F | p | df | F | p |
| System 1 | PC1 | 1.98 | 23.61 | <0.0001 | 1.99 | 40.05 | <0.0001 | 1 | 0.11 | 0.74 | 1 | 4.27 | 0.039 |
| | PC2 | 1 | 0.99 | 0.32 | 1 | 19.32 | <0.0001 | 1 | 6.20 | 0.013 | 1 | 33.70 | <0.0001 |
| | PC3 | 1.82 | 3.92 | 0.013 | 1.82 | 4.83 | 0.045 | 1 | 1.73 | 0.19 | 1 | 1.99 | 0.16 |
| | PC4 | 1 | 0.05 | 0.83 | 1 | 0.70 | 0.40 | 1 | 5.52 | 0.019 | 1 | 1.43 | 0.23 |
| | PC5 | 1.81 | 3.16 | 0.103 | 1 | 0.12 | 0.73 | 1 | 5.08 | 0.024 | 1 | 0.88 | 0.35 |
| | PCs | f(Temperature) × PCs | | | f(Salinity) × PCs | | | Chl-a × PCs | | | | | |
| | | edf | F | p | edf | F | p | df | F | p | | | |
| System 2 | PC1 | 1 | 4.39 | 0.04 | 1.99 | 53.66 | <0.0001 | 1 | 3.51 | 0.062 | | | |
| | PC2 | 1.98 | 31.33 | <0.0001 | 1 | 91.21 | <0.0001 | 1 | 3.03 | 0.082 | | | |
| | PC3 | 1 | 0.001 | 0.97 | 1 | 4.18 | 0.041 | 1 | 0.17 | 0.68 | | | |
| | PC4 | 1 | 1.17 | 0.28 | 1 | 1.80 | 0.18 | 1 | 0.001 | 0.97 | | | |
| | PC5 | 1.97 | 18.98 | <0.0001 | 1.86 | 10.98 | 0.00061 | 1 | 2.61 | 0.11 | | | |
| | PCs | f(enviro-PC1) × PCs | | | | | | | | | | | |
| | | edf | F | p | | | | | | | | | |
| System 3 | PC1 | 1.97 | 32.23 | <0.0001 | | | | | | | | | |
| | PC2 | 1.72 | 7.52 | 0.012 | | | | | | | | | |
| | PC3 | 1.48 | 1.82 | 0.10 | | | | | | | | | |
| | PC4 | 1.85 | 2.59 | 0.048 | | | | | | | | | |
| | PC5 | 1.94 | 15.12 | <0.0001 | | | | | | | | | |

**Table 2.** GAMM summary results for individual smooth and linear terms (System 1, 2 and 3). Estimated degrees of freedom, *F* statistics and significance values for each term from the interactions between environmental covariates and PCs are reported for individual study system.

In this study, the combination of EFA, GAMMs and multiple systems on different spatial scales allowed us to identify shell features under control of age and species factors, and uncouple these from the modelled shape variance to describe independent general and more local relationships between Atlantic *Mytilus* shape and natural environments.

**Quantifying environmental effects on shell shape.** Environmental influence on mussels is complex, with interacting factors that may result in a variety of shape patterns[27]. These interactions make it problematic to isolate effects of single drivers in natural environments and constrain predictive power[17,60].

Among these drivers, population density and predation influence responses in blue mussels including changes in shell proportions[28] and structure[61,62]. Genotypic differences and hybridisation within the *Mytilus edulis* species-complex are also known to influence spatial patterns of shell variation[6,10,12]. Moreover, although we considered the annual mean of environmental parameters, other factors could have substantial effects on mussel growth and shape, such as seasonality and the environmental conditions during specific life-stages[23,63]. However, it is not always possible to include all the interacting drivers at the different scales of analysis. We overcame these limitations through the study of blue mussel populations collected from systems on various geographical scales and with known genetic status, overall providing different degrees of control on regional confounding factors.

Specifically, in the aquaculture system (small-scale, System 3) the cultivation technique considerably reduces accessibility of predators[64] and densities are often actively controlled[65]. In mussel farms on Scottish west coast, multiple *Mytilus* species and hybrids generally occur in low frequencies and are geographically restricted[64]. Therefore, we used cultured mussels to identify fine-scale shape responses to different environmental exposures (depending on cultivation depth) in a habitat offering ideal conditions for rapid growth.

In the Baltic Sea (medium-scale, System 2), mussels constitute 80–90% of the coastal animal biomass[66] and have a strong advantage over competitors for space[67]. This dominance is attributed to an almost complete absence of predators[66,68]. Here, an increasing *M. edulis* × *M. trossulus* hybridisation with decreasing salinity has been reported[10,40]. Overall, this region offered low competition and predation across different salinities ranging from marine (~32 psu, outer basin) to brackish waters (~6 psu, inner basin).

Conversely, Northern Atlantic and Arctic *Mytilus* populations (large-scale, System 1) face variable predation pressures with latitude[69] and competition for space[28,33]. These confounding factors are generally difficult to quantify directly across a wide geographical scale. However, the possibility to demonstrate broad-scale shape patterns across latitudinal gradients and to compare these variations with more local trends (System 2 and 3) provided complementary information on the factors regulating mussel form and make it possible to draw more general conclusions on shell shape plasticity.

**Local shape variation.** In System 1 (Fig. 4a, Supplementary Figure S8), we observed a strong environmental influence on the shape variation captured by PC1, and additional temperature and chl-a effects on PC3 and PC4, respectively. According to documented growth trends of *Mytilus*[28], PC2 indicated a strong developmental (age) effect on shape variance, describing differences between young (round) mussels and old (curved with wider shells) individuals. Exponential variance structures indicated formation of more heterogeneous average shapes with decreasing temperature (Fig. 4d).

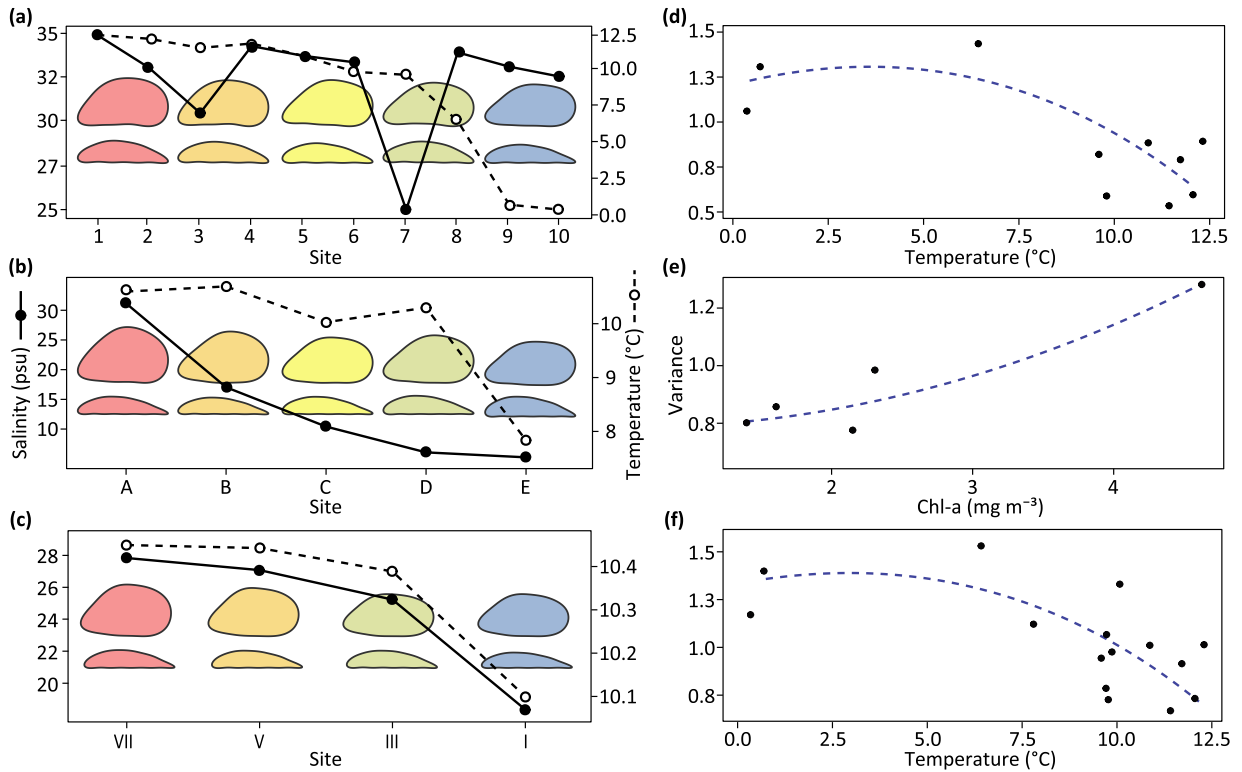

**Figure 4.** Response of *Mytilus* mean shape and heterogeneity to environmental variation. Left: graphs show a marked convergence of average shell shapes for the individual study systems. (**a**) System 1, large-scale North Atlantic and Arctic populations (only populations 1, 4, 5, 9 and 10 are shown in the background), (**b**) System 2, medium-scale Baltic region and (**c**) System 3, small-scale Loch Leven. The convergence of mean shell outlines at different geographical scales indicates the consistent formation of elongated, narrow shells and more parallel dorsoventral margins under lower temperature and salinity. Right: graphs represent system-wise patterns of shape heterogeneity with habitat conditions estimated from optimal variance structures within individual GAMMs. Loess smoothers (dashed lines) are fitted for visual interpretation. (**d**) The range of shell variation in North Atlantic and Arctic populations (System 1) showing the formation of more heterogeneous shapes in colder waters. (**e**) Positive trend of shape variance in the Baltic region (System 2) depicting more heterogeneous shell shapes with increasing food levels. (**f**) Pattern of shape variance for the Atlantic system showing an increase of shape heterogeneity with decreasing water temperature.

Mussels in System 2 (Fig. 4b, Supplementary Figure S9a) showed marked environmental effects on the shape captured by PC1 and PC2, although no correlation between salinity gradients and shell traits was found previously using traditional morphometrics[70]. We observed increasing elongation and shell width with decreasing salinity, indicating a stronger effect of this factor on shape than increasing mussel densities in the inner Baltic as previously thought[71]. The formation of more heterogeneous shapes for higher chl-a concentrations (Fig. 4e) suggests a strong effect of food availability on mussels growing at low salinities, especially near Baltic coastal lagoons, where food-enriched water inputs are markedly seasonal[72].

In System 3, we detected a strong environmental effect on shell shape (PC1-PC2; Fig. 4c, Supplementary Figure S9b). This highlights how altered growth rates[27] as well as decreasing stocking densities with depth (Michalek, pers. obs.) are likely to contribute to the shape variations along the cultivation rope.

Overall, we observed similar shell shape patterns at different geographical scales, consisting in the formation of elongated, narrow shells and more parallel dorsoventral margins with decreasing temperature, salinity and food supply. There was also a consistent overlap among PCs for the different study systems, except for PC2 from System 1, describing a strong age effect on mussel shapes due to the wide range of size classes available.

**General shape variation.** The optimal model for the Atlantic system showed more general environmental effects on shell shape and confirmed some of the detected local relationships (Fig. 3a). As with PC2 from System 1, PC1 revealed a strong age contribution. We detected a marked environmental influence on PC2, PC3 and PC4, demonstrating a strong effect of salinity on the shape responses in *Mytilus* (Fig. 3b). The absence of environmental or age effects on PC5 indicated a genetic (species) influence on the captured shape variance. We also detected an increased shape heterogeneity in colder waters corroborating the documented variance structure in System 1 (Fig. 4f).

| PCs | f(Temperature) × PCs | | | f(Salinity) × PCs | | | f(Chl-a) × PCs | | | Length × PCs | | |
|---|---|---|---|---|---|---|---|---|---|---|---|---|
| | edf | F | p | edf | F | p | edf | F | p | df | F | p |
| PC1 | 1.92 | 13.89 | <0.0001 | 2.00 | 101.32 | <0.0001 | 1.94 | 11.79 | <0.0001 | 1 | 70.64 | <0.0001 |
| PC2 | 1.98 | 26.67 | <0.0001 | 1 | 13.13 | 0.0003 | 1.88 | 5.77 | 0.0056 | 1 | 13.60 | 0.0002 |
| PC3 | 1.88 | 6.38 | 0.0012 | 1.26 | 13.36 | 0.00015 | 1 | 1.57 | 0.21 | 1 | 18.75 | <0.0001 |
| PC4 | 1 | 11.38 | 0.00076 | 1.98 | 23.71 | <0.0001 | 1.90 | 9.75 | <0.0001 | 1 | 1.74 | 0.19 |
| PC5 | 1.87 | 2.95 | 0.038 | 1.86 | 3.65 | 0.059 | 1.70 | 1.79 | 0.27 | 1 | 3.80 | 0.051 |
| | Estimate | 95%CI | | Estimate | 95%CI | | Estimate | 95%CI | | Estimate | 95%CI | |
| PC1 | −1.11 | −1.722; −0.500 | | 0.30 | −0.161; 0.756 | | 0.99 | 0.351; 1.631 | | 0.55 | 0.406; 0.696 | |
| PC2 | 0.40 | −0.171; 0.975 | | 0.82 | 0.382; 1.268 | | 0.53 | −0.095; 1.152 | | 0.26 | 0.119; 0.406 | |
| PC3 | −0.51 | −0.908; −0.120 | | −0.95 | −1.378; −0.514 | | 0.39 | −0.228; 1.016 | | −0.33 | −0.474; −0.190 | |
| PC4 | −1.12 | −1.716; −0.527 | | 0.83 | 0.389; 1.271 | | 1.34 | 0.702; 1.984 | | −0.10 | −0.244; 0.043 | |
| PC5 | −0.12 | −0.691; 0.444 | | −0.28 | −0.736; 0.173 | | 0.22 | −0.425; 0.855 | | −0.17 | −0.312; −0.022 | |

**Table 3.** GAMM summary results for individual smooth and linear terms (Equation 1). Estimated degrees of freedom, F statistics, significance values (upper table), mean effect size of predictors for each response variable (PCs) and 95%CIs (lower table), for each term from the interactions between environmental covariates and PCs are reported.

Exponential variance structures revealed new patterns of shape heterogeneity depending on the spatial scale analysed (Fig. 4d–f). The local trend observed for System 2 should be considered more of an independent case, showing how the strong salinity effect can be altered locally by increased food supply. On larger geographical scales, however, temperature had a stronger effect on shape heterogeneity. We observed heterogeneous mussel responses in colder waters, creating generally less favourable conditions for mussel growth[27,73]. Specifically, identified growth alterations might be more evident due to potential competitive advantages of some individuals under environments selecting for specific shapes. On the other hand, in warmer waters, among-individual shape differences may be less conspicuous due to generally more favourable conditions and higher growth potential[27].

**Trends in shell shape.** An analysis of the Atlantic system and the comparative use of smaller-scale study systems allowed us to identify general patterns of shape variation as well as independent local trends. Few differences were detected between the individual levels of analysis, while the explained shape features and associations were generally consistent.

The definition of independent variables (PCs) allowed the uncoupling of the individual components of shape variance affected by environmental, developmental (age) and genetic (species) factors, and the identification of the shell features characterised by the strongest shape alterations (Fig. 5). Specifically, PC1 captured a significant proportion of shape variance related to age modifications of shell outlines during growth (Fig. 5a). PC2 expressed the largest component of the environmental contribution to shape, describing shell variations under a marked salinity effect (Fig. 5b). Additional environmental contributions were described by PC3 and PC4 affected by temperature, salinity and strongly by food availability (Fig. 5c,d). PC5 (+ PC7, PC9, PC10) described the shape variance controlled by species identity and, therefore, the shell features discriminating between *M. edulis-* and *M. trossulus*-like specimens (Fig. 5e).

Overall, environmental variation influenced a larger proportion of shape variance (PC2+ PC3+ PC4: 49%) and exerted a stronger effect than age (PC1: 38%) and species identity (PC5+ PC7+ PC9+ PC10: 7%) on the shell shape variation in Atlantic *Mytilus*.

We detected similar *Mytilus* shape responses to less favourable conditions at the different scales of analysis, indicating the formation of elongated and narrow shells, with more parallel dorsoventral margins (Fig. 4). These variations could be explained by shapes being driven by the maintenance principle of a physiologically favourable surface-area to volume ratio[27], which increases in elongated shells. The observed shapes, along with physiological acclimatisation[18], could represent an important component of mussel adaptation to environmental stressors.

GAMMs demonstrated water salinity to have a stronger influence (effect size) than other predictors on mussel shape variation than previously reported[12,26,70]. Results suggest this physical parameter can lead to dramatic shape changes under sub-optimal conditions to cope with increased metabolic costs resulting from osmotic stress in low saline waters[25]. Our models also identified previously unrecognised *Mytilus* shape patterns revealing the formation of less heterogeneous outlines with increasing water temperature as well as more local effects of food supply on the variability of shape responses.

Our findings illustrate that the use of powerful statistical tools can improve our understanding of blue mussel ecology and allow more accurate predictions of shell shape variations in multivariate natural environments over space and time, with the potential application of these methods to a wide range of fields.

Climatic changes are projected to impact all areas of the ocean and different geographical regions are predicted to change at different rates[74]. Therefore, understanding the links between organismal responses and changing environmental conditions is essential if we want to accurately forecast spatial sensitivity patterns of foundation species, such as mussels, and their potential impacts on higher levels of organisation[17,22]. Indeed, disentangling the different factors affecting mussel shape can help to predict future changes in shell integrity and vulnerability to predators, with potential consequences from both an economic and ecological perspective[24,56].

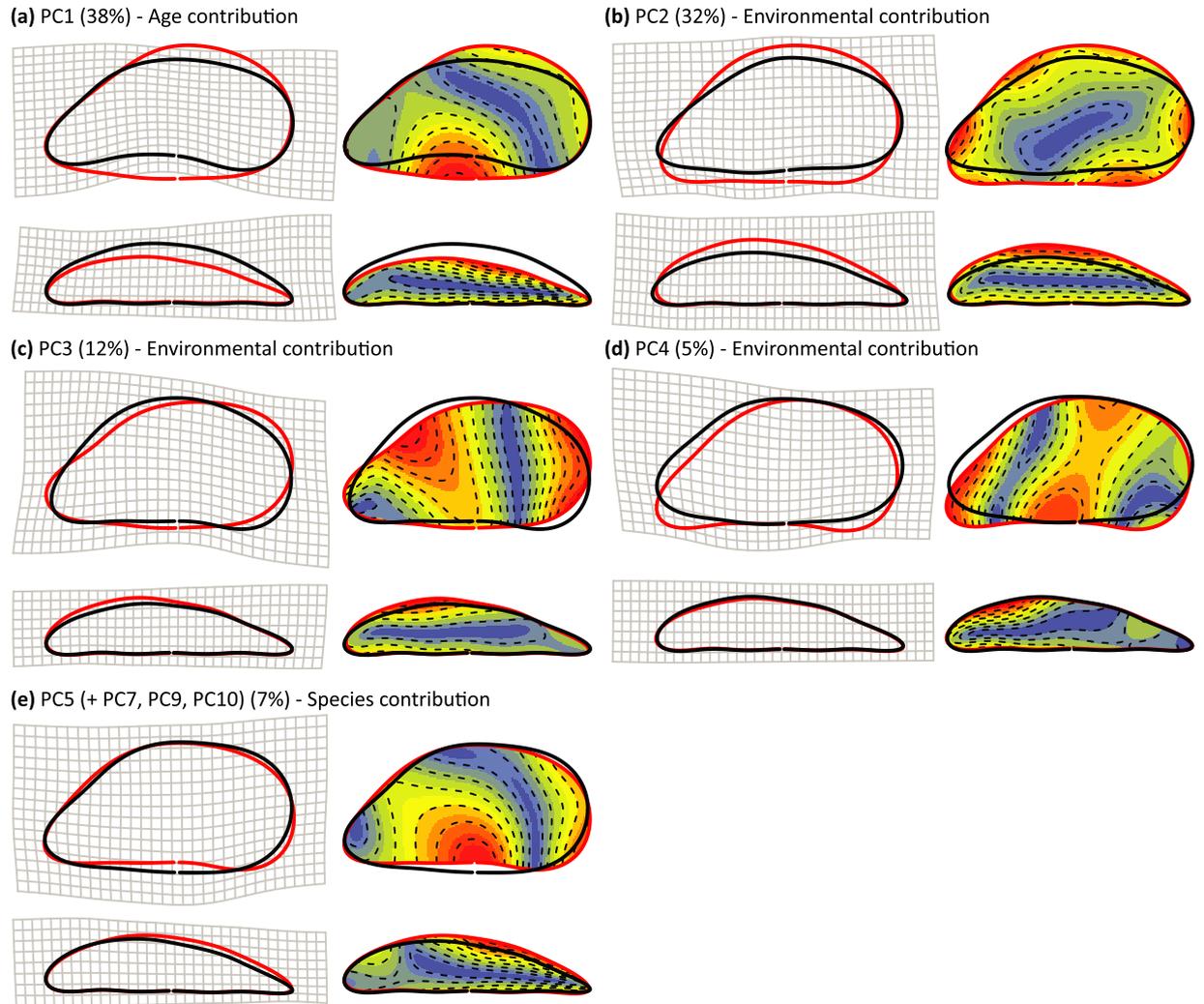

**Figure 5.** Effects of shape variance components on the outline reconstruction. Contributions to mean shapes for the individual components of shape variance regulated by age, environmental and species factors. The influence of each PC on the average *Mytilus* shapes, for both lateral and ventral views, were visualised with: deformation grids (left), depicting the bindings required to pass from the average shape for low (black) to high (red) PC values, and iso-deformation lines (right), representing the outline regions subject to different degrees of change (continuous scale from blue, low deformation, to red, strong deformation). (**a**) PC1 (38%, age contribution) showed a progressive elongation of the shell, with the formation of convex ventral margins, giving a curved aspect to the outline, and an increase of shell width with increasing shell length[28]. (**b**) PC2 (32%, environmental contribution) illustrated the formation of elongated and narrow shells with decreasing salinity. (**c**) PC3 (12%, environmental contribution) explained a progressive rounding of dorsal margins and concaving of ventral profiles for increasing temperature and salinity. (**d**) PC4 (5%, environmental contribution) reported an increase of ventral concaveness with decreasing food availability. (**e**) PC5 (+PC7, 9, 10) (7%, species contribution) indicated the development of concave ventral sides and round dorsal margins in *M. edulis*-like individuals (red) and elongated shells with parallel dorsoventral margins in *M. trossulus*-like specimens (black).

Our method allows the identification of specific environmental effects on shape variation and the use of contemporary mussels to forecast the responses of benthic communities to near-future climate changes at different regional scales. These findings could also be applied to the study of population responses with temporal clines and to promote the use of shell shape from fossils bivalves to understand past climates and environments.

## Conclusions

The combined use of EFA and GAMMs, and the employment of multiple study systems on a wide geographical scale made it possible to describe general relationships between shell shape variation in Atlantic *Mytilus* species and environmental drivers that are independent of developmental (age) and genetic (species) contributions to mussel shape. New methods allowed the identification of previously unrecognised patterns of mussel form and variations in specific shell features at a much finer scale than possible previously.

1. EFA of outlines on blue mussel populations covering a wide latitudinal range (28°, about 3980 km) allowed an in-depth quantification of shell shape and the definition of new independent variables expressing shape variations at different regional scales.
2. GAMMs and multiple levels of analysis (from small to large geographical scale) described general patterns as well as more local trends of natural shell shape variation and heterogeneity in blue mussels.
3. Powerful statistical methods allowed the identification of shell features under control of environmental, age and species (*M. edulis* and *M. trossulus*) factors. The ability to uncouple these individual components from the modelled shape variance made it possible to describe independent relationships between blue mussel shape and environment.
4. Models demonstrated that salinity has the strongest effect on the spatial patterns of shell shape variation, while temperature and food supply are the main drivers of shape heterogeneity, predicting potentially dramatic shape modifications in blue mussels under future environmental conditions.
5. Blue mussels showed similar shell shape responses to less favourable environmental conditions at different geographical scales, with the formation of elongated, narrow shells and more parallel dorsoventral margins, suggesting shell shape variability represents an important adaptive component to environmental stressors.

Although relationships between mussel shape and environmental factors were identified, more studies are needed to understand the adaptive significance of the observed alterations and their underlying causes. Therefore, by providing appropriate study systems and accurate ways to quantify animal shape and diversity, morphological variation can represent a powerful indicator for understanding the adaptation of organisms and help to predict their responses in a rapidly changing environment.

**Data Availability.** The datasets analysed and codes supporting this study are made available through the University of Cambridge data repository and can be accessed at https://doi.org/10.17863/CAM.12536.

### Acknowledgements
We thank the three anonymous reviewers whose comments and suggestions helped to clarify and substantially improve this manuscript. We are also grateful to Iain Johnston (Scottish Oceans Institute, St Andrews, UK), Dr. Coleen Suckling (School of Ocean Sciences, Bangor, UK), Sarah Dashfield (Plymouth Marine Laboratory, Plymouth, UK), Dr. Peter Thor (Norwegian Polar Institute, Tromsø, Norway), Alexander Ventura (University of Gothenburg, Kristineberg, Sweden), Kirti Ramesh (GEOMAR, Kiel, Germany), Dr. Henk van der Veer and Rob Dekker (Royal Netherlands Institute for Sea Research, Texel, Netherlands) for their help with mussel specimen collection, the Statistics Clinic (University of Cambridge, UK) for useful statistical advice and Dr. Vincent Bonhomme (Institut des Sciences de l'Évolution, Montpellier, France) for help on the use of the Momocs package. The work was funded by the European Union Seventh Framework Programme, Marie Curie ITN under grant agreement n° 605051.

### Author Contributions
L.T., L.P., E.M.H. conceived the original project and designed the study; L.T. carried out the data acquisition, morphometric and statistical analyses; L.T., K.M., T.S. and J.T. collected material for the study; K.M. provided environmental monitoring data for cultured mussels; L.T., L.P., E.M.H. wrote the manuscript. All authors edited the final manuscript.

### Additional Information
**Supplementary information** accompanies this paper at https://doi.org/10.1038/s41598-018-20122-9.

**Competing Interests:** The authors declare no competing interests.

**Publisher's note:** Springer Nature remains neutral with regard to jurisdictional claims in published maps and institutional affiliations.